\documentclass[letterpaper]{article} 
\usepackage{aaai24}  
\usepackage{times}  
\usepackage{helvet}  
\usepackage{courier}  
\usepackage[hyphens]{url}  
\usepackage{graphicx} 
\usepackage{natbib}  
\usepackage{caption} 
\usepackage{algorithm}
\usepackage{algorithmic}
\usepackage{amssymb}
\usepackage{amsmath}  
\usepackage{multirow}
\usepackage{booktabs}
\usepackage{subfigure}
\usepackage{newfloat}
\usepackage{listings}
\DeclareCaptionStyle{ruled}{labelfont=normalfont,labelsep=colon,strut=off} 
\floatstyle{ruled}
\newfloat{listing}{tb}{lst}{}
\floatname{listing}{Listing}
\title{Ada-Retrieval: An Adaptive Multi-Round Retrieval Paradigm for Sequential Recommendations}
\author {
    Lei Li\textsuperscript{\rm 1},
    Jianxun Lian\textsuperscript{\rm 2},
    Xiao Zhou\textsuperscript{\rm 1}\thanks{Corresponding Author.},
    Xing Xie\textsuperscript{\rm 2}
}
\affiliations {
    \textsuperscript{\rm 1}Gaoling School of Artificial Intelligence, Renmin University of China\\
    \textsuperscript{\rm 2}Microsoft Research Asia\\
    leil@ruc.edu.cn, jianxun.lian@outlook.com, xiaozhou@ruc.edu.cn, xingx@microsoft.com
}

\usepackage{bibentry}

\begin{document}

\maketitle

\begin{abstract}
Retrieval models aim at selecting a small set of item candidates which match the preference of a given user. They play a vital role in large-scale recommender systems since subsequent models such as rankers highly depend on the quality of item candidates. However, most existing retrieval models employ a single-round inference paradigm, which may not adequately capture the dynamic nature of user preferences and stuck in one area in the item space. In this paper, we propose Ada-Retrieval, an adaptive multi-round retrieval paradigm for recommender systems that iteratively refines user representations to better capture potential candidates in the full item space. Ada-Retrieval comprises two key modules: the item representation adapter and the user representation adapter, designed to inject context information into items' and users' representations. The framework maintains a model-agnostic design, allowing seamless integration with various backbone models such as RNNs or Transformers. We perform experiments on three widely used public datasets, incorporating five powerful sequential recommenders as backbone models. Our results demonstrate that Ada-Retrieval significantly enhances the performance of various base models, with consistent improvements observed across different datasets. Our code and data are publicly available at: https://github.com/ll0ruc/Ada-Retrieval.
\end{abstract}

\section{Introduction}
Recommender systems have become a crucial element in a wide range of online applications, encompassing e-commerce, social media, and entertainment platforms~\cite{covington2016deep,ying2018graph}. By providing personalized recommendations tailored to users' historical behavior and preferences, these systems enhance user experience and engagement. Among the diverse types of recommender systems, sequential recommender systems~\cite{rendle2010factorization,tang2018personalized} have attracted considerable interest due to their capacity to effectively capture temporal dynamics in user history and accurately forecast near-future user behaviors. In this domain, various backbone models have been proposed, including recurrent neural networks (RNNs)~\cite{hidasi2015session}, convolutional neural networks (CNNs)~\cite{tang2018personalized}, transformers~\cite{kang2018self}, and graph neural networks (GNNs)~\cite{wu2019session}, each contributing to the ongoing advancement of sequential recommendation techniques.

\begin{figure}
    \centering
    \subfigure[Single-round retrieval]{\includegraphics[width=0.22\textwidth]{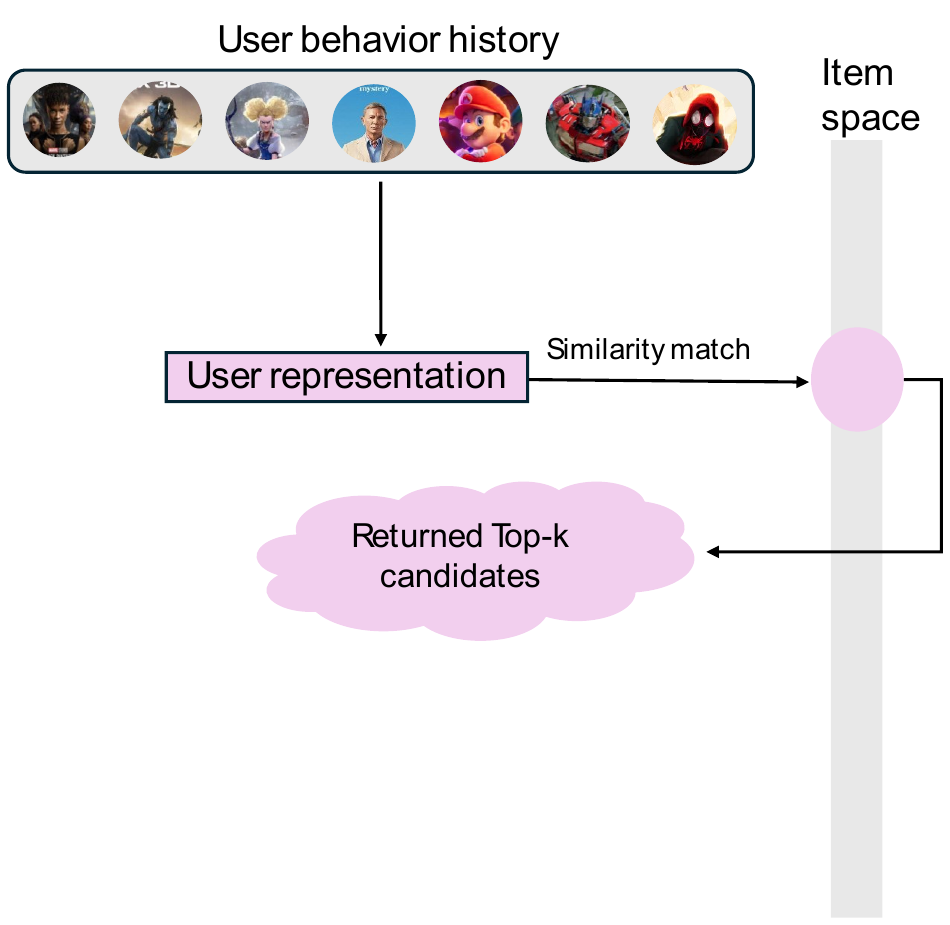} } 
    \subfigure[Multi-round retrieval]{\includegraphics[width=0.22\textwidth]{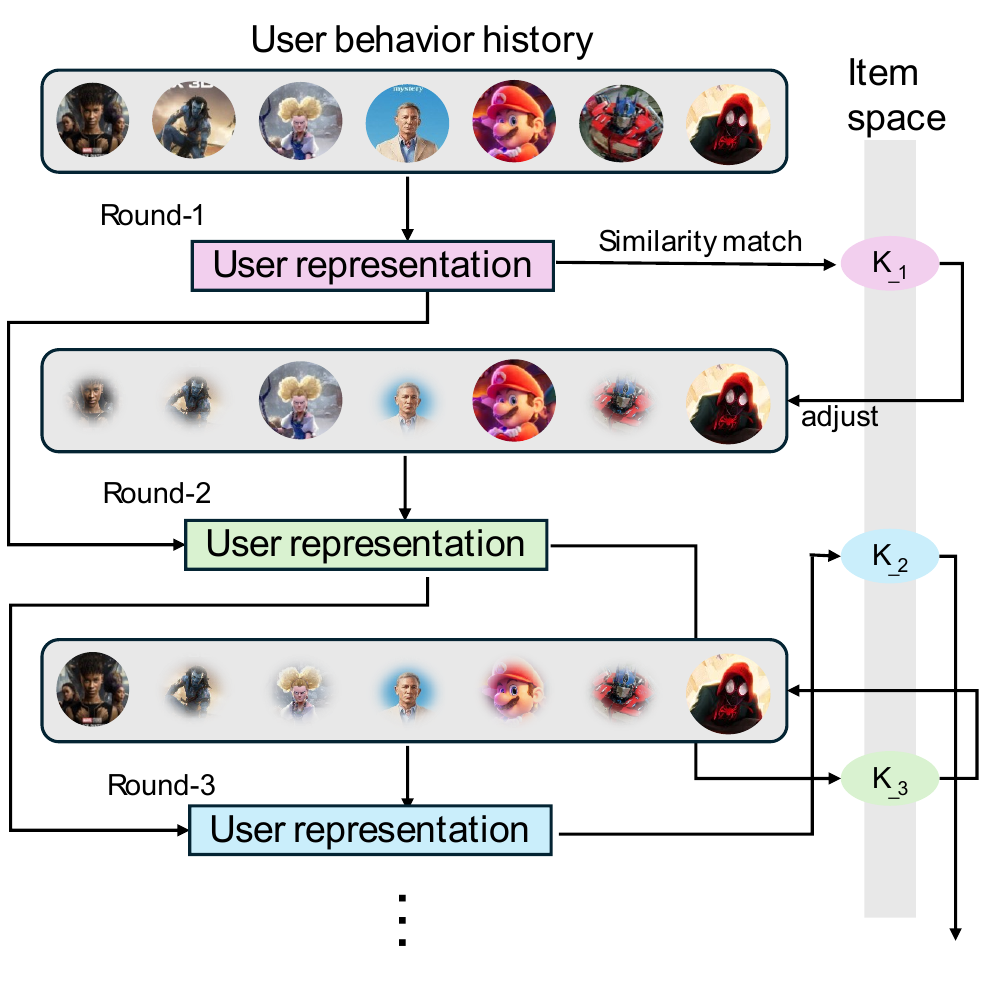}} 
    \caption{Illustrations of (a) the conventional single-round retrieval paradigm, and (b) our proposed adaptive multi-round retrieval paradigm, in which the final retrieval result is the union of each individual retrieval $K_i$.}
    \label{fig:intro}
\end{figure}

This paper does not aim to propose a stronger backbone model. Instead, we observe that most existing models employ a \textbf{single-round inference} paradigm to retrieve the top-$k$ item candidates~\cite{he2016fusing,tang2018personalized,sun2019bert4rec}. Specifically, given users' profiles such as behavior histories, the model initiates the forward process and generates user representations, which are then used as queries to match the top-$k$ most similar items in the database. However, this single-round inference paradigm may not adequately capture the dynamic nature of user preferences and adapt to the ever-changing diversity of the item space. As illustrated in Figure~\ref{fig:intro}(a), once the model's forward pass is completed, the user representation remains fixed, resulting in a top-$k$ search area in the item space that is confined to a static region. If the initial user representation is inaccurate or the user's future preferences are diverse, this paradigm may fail to deliver satisfactory performance.

We argue that a \textbf{multi-round inference} paradigm offers a more effective retrieval approach for recommender systems. As illustrated in Figure~\ref{fig:intro}(b), the objective of retrieving $k$ items is divided into $n$ batches, with each batch representing a round of retrieving $k/n$ items. The forward passes of user representation in different rounds are conducted independently. If the previously retrieved items do not adequately match the user's preferences, the user representation will be adjusted in the next round, allowing the model to search for target items in a different region of the item space. Taking the search engine scenario as an analogy~\cite{zhang2006mining}, users may rewrite their queries if the currently retrieved information does not accurately address their questions. In this regard, previous rounds' retrieval can serve as feedback information~\cite{lewandowski2008retrieval}, which helps refine the user representation if necessary. Thus, this multi-round paradigm presents a significant advantage, as it prevents user representations from being confined to a static area, enabling more dynamic and diverse recommendations.

As an embodiment of the new paradigm, we present Ada-Retrieval, an adaptive multi-round retrieval approach for recommender systems. Fundamentally, Ada-Retrieval alters the traditional training and inference process while maintaining a \textbf{model-agnostic design}, allowing seamless integration with various backbone models such as RNNs or Transformers. Ada-Retrieval comprises two key modules: the item representation adapter and the user representation adapter. Both modules aim to inject context information, which refers to previous user representations and retrieved items up to the current retrieval round, into items' and users' representations. The item representation adapter consists of a learnable filter (LFT) layer and a context-aware attention (CAT) layer, designed to adjust item embeddings in the user history according to the retrieval context. This enables the user model to potentially optimize for the next round of retrieval by considering the feedback of item candidate space. The user representation adapter, on the other hand, is composed of a Gated Recurrent Units (GRU) layer and a Multi-Layer Perceptron (MLP) layer. The GRU layer encodes all user representations generated in previous rounds as user context, while the MLP layer fuses this context with the current user representation to produce an adapted one. By incorporating these components, Ada-Retrieval can integrate contextual information during the retrieval process into traditional sequential recommendation models, generating progressively refined user representations for item retrieval while maintaining a lightweight and model-agnostic advantage.

We perform experiments on three widely used public datasets, incorporating five powerful sequential recommenders as backbone models. Comprehensive results demonstrate that Ada-Retrieval can significantly enhance the performance of various base models, with consistent improvements observed across different datasets. For instance, on the Beauty dataset, Ada-Retrieval boosts SASRec's performance by 8.55\% in terms of NDCG@50 and improves the best base model, FMLPRec, by 5.66\%. The key contributions of this paper are summarized as follows:

\begin{itemize}

\item We propose Ada-Retrieval, a novel adaptive multi-round retrieval framework for sequential recommendations. Unlike traditional single-round retrieval, Ada-Retrieval iteratively refines user representations to better capture potential candidates across the entire item space.

\item We design several key components, including LFT and CAT for the item representation adapter, and GRU and MLP for the user representation adapter. These components enable the integration of contextual information in a model-agnostic manner.

\item We conduct extensive experiments on real-world datasets to demonstrate the effectiveness of Ada-Retrieval, showing significant improvements over various sequential recommender systems.
\end{itemize}

\section{Related Work}

\subsection{Deep Retrieval}
In practical recommender systems, the retrieval stage (candidate generation) aims to efficiently retrieve a small subset of items, typically in the hundreds, from large corpora~\cite{xie2020internal}. With the rise of deep learning, there has been a surge in efforts to construct sophisticated retrieval models for recommender systems. Embedding-based methods often adopt a two-tower architecture, as seen in FM~\cite{rendle2010factorization}, YoutubeDNN~\cite{covington2016deep}, and AFT~\cite{hao2021adversarial}, dividing the construction of user and item representations into two distinct branches. Innovations like TDM~\cite{zhu2018learning} and JTM~\cite{zhu2019joint} offer fresh perspectives on leveraging user-item dynamics through tree-based structures. Additionally, graph-based matching models~\cite{xie2021improving} are proposed to learn user/item representations. Departing from the single-round inference paradigms of these methods, our model introduces a multi-round inference paradigm, providing a more effective retrieval approach for recommender systems.

\subsection{Sequential Recommendation}
Sequential recommendation, predicting future items to interact with based on correlations in item transitions within user activity sequences, has evolved from foundational Markov Chain models~\cite{he2016fusing} to contemporary deep learning technologies. Caser~\cite{tang2018personalized} employed CNNs to analyze sequences of item embeddings, while GRU4Rec~\cite{hidasi2015session} used Gated Recurrent Units (GRU) for session-driven recommendations. More recently, SASRec~\cite{kang2018self} incorporated self-attention mechanisms to selectively aggregate relevant items, refining user modeling. Inspired by the Cloze task, Bert4Rec~\cite{sun2019bert4rec} predicted masked items by jointly utilizing preceding and succeeding contexts. Training frameworks like CL4SRec~\cite{xie2022contrastive} integrated contrastive approaches for diverse perspectives through data enhancement. Despite the success of these models, a challenge remains in generating diverse user feature representations. In addressing this, our model iteratively refines user representations, enhancing the capture of dynamic user preferences through inserted contextual information.

\begin{figure*}[t]
\centering
\includegraphics[width=\textwidth]{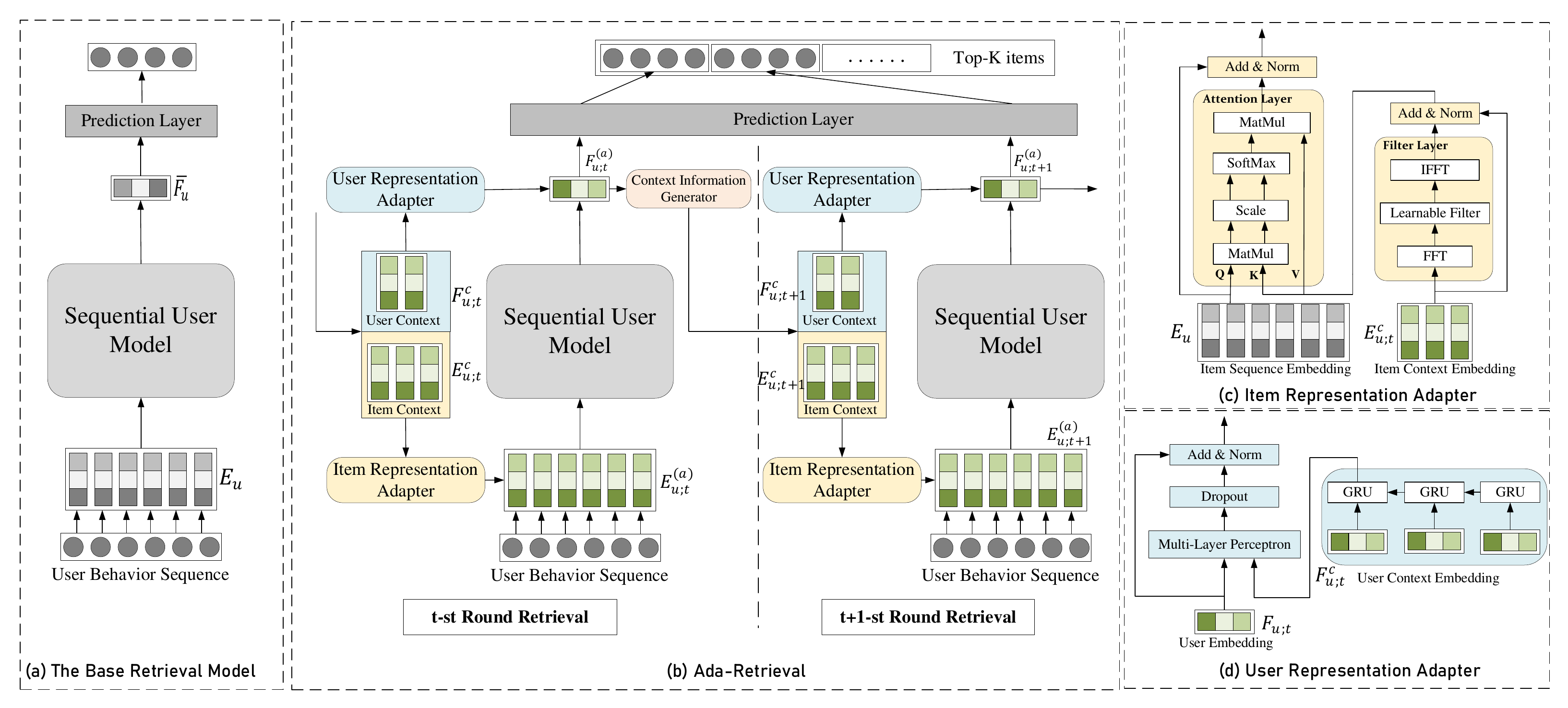} 
\caption{An overview of the traditional retrieval model (a) and our proposed Ada-Retrieval paradigm (b), which consists of two key parts: the item representation adapter (c) and the user representation adapter (d). We use colored elements to indicate the new components in Ada-Retrieval.}
\label{fig:model}
\end{figure*}

\section{Preliminaries}
\subsection{Problem Formulation}
Let us assume a set of users $\mathcal{U} = \{u_1,u_2,\dots, u_{|\mathcal{U}|}\}$ and items $\mathcal{I} = \{i_1,i_2,\dots, i_{|\mathcal{I}|}\}$, with $u \in \mathcal{U}$ representing a user and $i \in \mathcal{I}$ representing an item. The user behavior can be denoted as $\mathcal{S} = \{s_1,s_2,\dots, s_{|\mathcal{U}|}\}$. In sequential recommendation, a user's behavior sequence is typically ordered chronologically: $s_u = \{i_1,i_2,\dots, i_n\}$, where $s_u \in \mathcal{S}$ and $u \in \mathcal{U}$. The objective of sequential recommendation is to predict the next item the user is likely to interact with, denoted as $p(i_{n+1}|i_{1:n})$.

\subsection{Base Sequential Model}
A common architecture for a sequential recommender system typically consists of three key components: an embedding lookup layer $\rm EMB(\cdot)$, a sequential encoding layer $\rm SEL(\cdot)$, and a prediction layer $\rm PRED(\cdot)$, as illustrated in Figure~\ref{fig:model}(a). When provided with a user behavior sequence $s_u = \{i_1,i_2,\dots, i_n\}$, the sequence initially passes through the embedding lookup layer $\rm EMB(\cdot)$, resulting in a sequence of corresponding item embeddings:
\begin{equation}
    \mathbf{E_u} = \rm EMB(s_u) = \{\mathbf{e_1},\mathbf{e_2},\dots,\mathbf{e_n}\}
\end{equation}
Subsequently, the embeddings of this item sequence are processed through a sequential user encoder, represented as $\rm SEL(\cdot)$, which can be implemented using appropriate backbones such as RNNs or Transformers:
\begin{equation}
	\mathbf{F_u} = \rm SEL(\mathbf{E_u})
\end{equation}
Here, $\mathbf{F_u}$ represents the embedding vector serving as the user representation. Then, $\mathbf{F_u}$ is combined with a target item vector $\mathbf{E_i}$ as the input to the prediction layer $\rm PRED(\cdot)$:
\begin{equation}
	\hat{y}_{ui} = \rm PRED(\mathbf{F_u}, \mathbf{E_i})
\end{equation}
The prediction layer is commonly implemented using either a dot product or cosine similarity, particularly for retrieval purposes.

\section{Methodology}
Our model, Ada-Retrieval, introduces an adaptive multi-round retrieval approach for recommender systems. The overall framework is depicted in Figure~\ref{fig:model}(b). At the core of the adaptive retrieval paradigm are two meticulously crafted adaptation modules: the Item Representation Adapter (IRA) and the User Representation Adapter (URA). These modules seamlessly integrate contextual information into user preference modeling. In contrast to traditional sequential recommendation models, our modifications primarily focus on the input and output of the sequential encoding layer $\rm SEL(\cdot)$:
\begin{equation}
\mathbf{E_u^{(a)}} = \rm IRA(\mathbf{E_u};\mathbf{E_u^c})
\end{equation}
\begin{equation}
\mathbf{F_u^{(a)}} = \rm URA(\mathbf{F_u};\mathbf{F_u^c})
\end{equation}
where $\mathbf{E_u^c}$, $\mathbf{F_u^c}$ are the feature representation of item context and user context. $\mathbf{E_u^{(a)}}$ and $\mathbf{F_u^{(a)}}$ represent the adjusted feature representation of item/user. In the next section, we will introduce the details of each proposed component.

\subsection{Item Representation Adapter}
The item representation adapter is designed to recalibrate item embeddings within users' historical data based on the prevailing retrieval item context. Further details are illustrated in Figure~\ref{fig:model}(c).

\subsubsection{Learnable Filter Layer}
Considering potential noise in item context information from previous rounds, we use a single learnable filter block for refining item features efficiently. This approach draws inspiration from the filter-enhanced MLP~\cite{zhou2022filter} used in recommendation systems, which typically employs multiple stacked blocks to enhance item feature representations by removing noise.

Upon receiving the item context information for the current round, denoted as $\mathcal{C}_u^t=\{i_1,i_2,\cdots, i_{\mathcal{C}_u^t}\}$, where $i$ represents items retrieved in previous rounds, our initial step involves passing it through an encoder layer to extract its features, $\mathbf{E_u^c} = \rm EMB(\mathcal{C}_u^t)$. Subsequently, we apply the Fast Fourier Transform (FFT), denoted as $\mathcal{F}(\cdot)$, along the item dimension. This operation transforms the item context representation matrix $\mathbf{E_u^c}$ into the frequency domain:
\begin{equation}
\mathbf{X_u^c} = \mathcal{F}(\mathbf{E_u^c})
\end{equation}
Note that $\mathbf{X_u^c}$ is a complex tensor representing the spectrum of $\mathbf{E_u^c}$. We can then proceed by multiplying it with a learnable filter, denoted as $W$:
\begin{equation}
\mathbf{\widetilde{X}_u^c} = W \odot \mathbf{X_u^c}
\end{equation}
where $\odot$ is the element-wise multiplication. Finally, we employ the inverse FFT to revert the modulated spectrum, $\mathbf{\widetilde{X}_u^c}$, back into the time domain, subsequently updating the sequence representations:
\begin{equation}
\mathbf{\widetilde{E}_u^c} = \mathcal{F}^{-1} (\mathbf{\widetilde{X}_u^c})
\end{equation}
where $\mathcal{F}^{-1}(\cdot)$ denotes the inverse 1D FFT, which converts the complex tensor into a real number tensor. To avoid overfitting, dropout layer~\cite{srivastava2014dropout}, residual connection structure~\cite{he2016deep}, and layer normalization operations~\cite{ba2016layer} are applied on the obtained output $\mathbf{H_u^c}$:
\begin{equation}
\mathbf{H_u^c} = \rm LayerNorm(\mathbf{E_u^c} + \rm Dropout(\mathbf{\widetilde{E}_u^c}))
\label{eq:pureitem}
\end{equation}

\subsubsection{Context-aware Attention Layer}
Attention mechanisms have proven effective in recommender systems~\cite{kang2018self, tan2021dynamic}. They empower the model to selectively focus on different segments of the sequence based on their relevance to the immediate prediction task. The following is the standard dot-product attention:
\begin{equation}
    \rm Attention(\mathbf{Q},\mathbf{K},\mathbf{V}) = \rm softmax(\frac{\mathbf{Q}\mathbf{K}^T}{\sqrt{d}})\mathbf{V}
\end{equation}
Here, $\mathbf{Q}$ denotes queries, $\mathbf{K}$ stands for keys, and $\mathbf{V}$ represents values. The embedding size is denoted by $d$, and the scale factor $\sqrt{d}$ is introduced to prevent excessively large values in the inner product.

In the typical self-attention mechanism, $\mathbf{Q}$, $\mathbf{K}$, and $\mathbf{V}$ are derived from the same input vector but are produced using distinct weight matrices. However, in our scenario, $\mathbf{Q}$ corresponds to the item sequence features, while $\mathbf{K}$ embodies the item context features. Therefore, the context-aware attention mechanism can be articulated as:
\begin{equation}
   \mathbf{\widetilde{H}_u^c} =\rm Attention(\mathbf{E_u},\mathbf{H_u^c},\mathbf{H_u^c}) 
\end{equation}
Certain pieces of item context information wield more influence in determining the subsequent item with which the user might engage. The attention mechanism enables the model to autonomously pinpoint these pivotal items, bestowing upon them greater weights. Following this, we incorporate the layer normalization and dropout operations to alleviate the gradient vanishing and unstable training problems as:
\begin{equation}
\mathbf{E_u^{(a)}} = \rm LayerNorm(\mathbf{E_u} + \rm Dropout(\mathbf{\widetilde{H}_u^c}))
\label{eq:norm}
\end{equation}

\subsubsection{A Case with Sequential User Model}
Following the acquisition of the adjusted item sequence feature representation, it seamlessly integrates into a conventional sequential recommendation model. Taking SASRec as an illustration, the user feature representation is denoted as $\mathbf{F_u} = \rm TRFM(\mathbf{E_u})$, where $\rm TRFM(\cdot)$ signifies the Transformer architecture within SASRec. In the context of Ada-Retrieval, the user feature representation is expressed as $\mathbf{F_u} = \rm TRFM(\mathbf{E_u^{(a)}})$, representing a modification to the input. It is noteworthy that Ada-Retrieval, with its model-agnostic nature, refrains from altering the intrinsic parameters of SASRec. Instead, it dynamically adjusts the current item sequence input features based on the item context information.

\subsection{User Representation Adapter}  
Utilizing the available user context information, we formulate the design of the user representation adapter to produce adaptive user representations, as illustrated in Figure~\ref{fig:model}(d).

\subsubsection{Gated Recurrent Unit Layer}
Recurrent Neural Networks (RNNs) have been developed to model variable-length sequence data~\cite{sherstinsky2020fundamentals}, showcasing promising advancements in recommendation systems~\cite{li2017neural, guo2020attentional}. Their efficacy stems from their capacity to capture a user's sequential behavior. Gated Recurrent Units (GRUs)~\cite{cho2014properties} represent a more sophisticated variant of RNNs designed to address the vanishing gradient challenge. Essentially, user features derived from earlier rounds exert influence on the current one, with this influence diminishing as the round distance grows. Therefore, we leverage the capabilities of the GRU module to encode user representations accumulated from previous rounds.

\begin{equation}
    \mathbf{\widetilde{F}_u^c} = \text{GRU}(\mathbf{F_u^c})
\label{eq:gru}
\end{equation}
$\mathbf{F_u^c}$ serves as the feature representation encapsulating user context, consisting of adjusted user feature representations generated in previous rounds.

With the trivial feature extractor of the user context, we essentially use the final hidden state as the representation of the user’s context representation $\mathbf{\widetilde{F}_u^c}$.

\subsubsection{Multi-Layer Perceptron Layer}
After deriving the user's context feature representation $\mathbf{\widetilde{F}_u^c}$, we concatenate it with the currently generated user feature representation $\mathbf{F_u}$. This combined representation then undergoes processing through a two-layer Multilayer Perceptron (MLP) with a ReLU activation function. The process is defined as follows:
\begin{equation}
    \mathbf{F_u^{(a)}} = W_2\, \text{ReLU}(W_1[\mathbf{\widetilde{F}_u^c};\mathbf{F_u}]+b_1)+b_2
\label{eq:ffn}
\end{equation}
where $W_1$, $b_1$, $W_2$, $b_2$ are trainable parameters. 

Subsequently, we incorporate skip connection and layer normalization operations, as detailed in Eq.~(\ref{eq:norm}), to produce the final user representation $\mathbf{F_u^{(a)}}$.

\subsection{Context Information Generator}
To collect the user context information generated in each round, we utilize a stacking methodology to assemble an array of context vectors:
\begin{equation}
    \mathbf{F_{u;t}^c} = \rm STACK(\{\mathbf{F_{u;1}^{(a)}},\mathbf{F_{u;2}^{(a)}},\cdots, \mathbf{F_{u;t-1}^{(a)}}\})
    \label{eq:usercontext}
\end{equation}

Concurrently, from the entire pool of candidate items, we retrieve the top-$k$ items with the highest scores aligned with the user representation of the current round, $\mathbf{F_u^{(a)}}$. The corresponding item IDs are then added to the item context pools:
\begin{equation}
    \mathcal{C}_u^t =  \mathcal{C}_u^{t-1} +  \text{top-}k \,\underset{i \in \mathcal{I}}{\rm argmax} \, \text{Sim}(\mathbf{F_{u;t-1}^{(a)}};\mathbf{E_i})
    \label{eq:itemcontext}
\end{equation}
where $\rm Sim$ is a function to measure the feature similarity between users and items. Here, we utilize the dot product for this purpose. It is crucial to highlight that the item context, determined through this similarity measure, will subsequently be input into the embedding look-up layer to retrieve their corresponding feature representations.

\subsection{Model Prediction and Optimization}
After $T$ iterative rounds, Ada-Retrieval yields $T$ user representations $\{\mathbf{F_{u;1}^{(a)}},\mathbf{F_{u;2}^{(a)}},\cdots,\mathbf{F_{u;T}^{(a)}} \}$. Then we multiply it by the item embedding matrix $\mathbf{E}$ to predict the relevance of the candidate item:
\begin{equation}
    \hat{y}_{ui;t} = \mathbf{E_i^T} \mathbf{F_{u;t}^{(a)}}
    \label{eq:dot}
\end{equation}
Each of these representations is employed to retrieve $k/T$ items, which are then sequentially concatenated to form the final set of top-$k$ items.

We anticipate that the actual item $i$ chosen by user $u$ should result in a higher score $\hat{y}_{ui}$. Hence, we utilize the cross-entropy loss to optimize the model parameters. The objective function for the $t$-th round is formulated as:
\begin{equation}
\small
    \mathcal{L}_t = -\sum_{u\in \mathcal{U},i \in \mathcal{I}}y_{ui;t}\log(\sigma(\hat{y}_{ui;t}))+(1-y_{ui;t})\log(1-\sigma(\hat{y}_{ui;t}))
\end{equation}
To emphasize early and accurate predictions of the positive item, we introduce a decay factor $\lambda$ to weigh each round's contribution to the overall training loss as:
\begin{equation}
    \mathcal{L} = \sum_{t=1}^T \lambda ^ t \mathcal{L}_t
    \label{eq:loss}
\end{equation}

To optimize training efficiency, we employ a two-phase training approach. Initially, we pre-train a foundational sequential model. Subsequently, in the second phase, we fine-tune Ada-Retrieval, utilizing the pre-trained model as a starting point. In this phase, we concurrently update both the parameters $\Theta$ of Ada-Retrieval and the $\Phi$ of the base model.

Ada-Retrieval is designed to seamlessly integrate with various retrieval models, including GRU-based and GNN-based models, preserving their structure while enhancing performance. However, it is not compatible with models like Matrix Factorization (MF) that rely on user-item interaction matrices.

\section{Experiments}
\subsection{Experimental Settings}

\subsubsection{Datasets.}
To validate our proposed method across diverse data types, we assess the model using three publicly available benchmark datasets. Beauty and Sports represent subsets of the Amazon Product dataset~\cite{mcauley2015image}, capturing user reviews of Amazon.com products. The Yelp dataset\footnote{https://www.yelp.com/dataset} is a sizable collection of extended item sequences derived from business recommendations, using only transaction records post-January 1st, 2019. For uniformity, we categorize interaction records by users or sessions and sequence them chronologically based on timestamps. Following~\cite{sun2019bert4rec, li2020time}, we filter out users/items with fewer than 5 interactions. Detailed statistics for each dataset are summarized in Table~\ref{tab:datasets}.

\begin{table}[!tb]
\centering
\begin{tabular}{l|ccc}
    \toprule
    Dataset  & Beauty & Sports & Yelp \\
    \midrule
    \# Sequences 	& 22,363 & 25,598 & 30,431  \\
    \# Items	& 12,101 & 18,357 & 20,033 \\
    \# Actions 	& 198,502 & 296,337 & 316,354  \\
    \# Sparsity 	& 99.93\% & 99.95\% & 99.95\%  \\
    \bottomrule
\end{tabular}
\caption{Statistics of datasets after preprocessing.}
\label{tab:datasets}
\end{table}

\subsubsection{Evaluation Settings.}
To facilitate comprehensive model evaluation, we employ the leave-one-out strategy~\cite{kang2018self,zhou2020s3} for partitioning each user's item sequence into training, validation, and test sets. Diverging from conventional sampling practices, our approach considers all items not previously engaged with by the user as candidate items~\cite{krichene2020sampled}. The evaluation metrics adopted for assessing model performance encompass top-\textit{k} Hit Ratio (HR@\textit{k}) and top-\textit{k} Normalized Discounted Cumulative Gain (NDCG@\textit{k}). 

\subsubsection{Implementation Details.}
When comparing with existing models, we adopt the optimal parameter settings from their original papers and conduct a meticulous grid search around these configurations for baseline models. Ada-Retrieval is implemented using Python 3.8 and PyTorch 1.12.1, executed on NVIDIA V100 GPUs with 32GB memory. Training parameters include an Adam optimizer with a learning rate of 0.001 and a batch size of 1024. Across all datasets, we set the maximum sequence length to 50, embedding dimension to 64, and training epochs to a maximum of 200. For Ada-Retrieval, we varied hyperparameters $T$ and $\lambda$ within the ranges [3, 8] and [0.1, 0.9], respectively, with step sizes of 1 and 0.2. The experiments were conducted five times, and results, reported as averages with standard deviations, reflect the model's performance. We also employed an early-stopping strategy, halting training if HR@50 performance on the validation set continuously decreased over 10 consecutive epochs.

\subsection{Main Results with Various Backbone Models}

\subsubsection{Backbones.}
As Ada-Retrieval is model-agnostic, we evaluate its performance with representative sequential recommenders, including GRU4Rec~\cite{hidasi2015session}, SASRec~\cite{kang2018self}, NextItNet~\cite{yuan2019simple}, SRGNN~\cite{wu2019session}, and FMLPRec~\cite{zhou2022filter}. These models employ diverse architectures, encompassing RNN, CNN, GNN, and MLP.

\subsubsection{Results.}

\begin{table*}[!t]
\centering
\scalebox{0.88}{
\begin{tabular}{c|l|cc|cc|cc|cc|cc} 
    \toprule
    \multicolumn{1}{c|}{\multirow{2} * {Datasets}} &\multicolumn{1}{c|}{\multirow{2} * {Models}}&\multicolumn{2}{|c}{GRU4Rec}&\multicolumn{2}{|c}{SASRec} &\multicolumn{2}{|c}{NextItNet}  &\multicolumn{2}{|c}{SRGNN} &\multicolumn{2}{|c}{FMLPRec}\\
    
    &  & HR & NDCG & HR & NDCG & HR & NDCG & HR & NDCG & HR & NDCG \\
    \midrule
    \multirow{3} * {Beauty}
     	 	          
    & Base & 13.126 & 4.574 & 17.110 & 6.506 & 12.539 &4.064  &12.411  &4.242  & 17.935 & 6.876  \\
    & Ada-Retrieval & 14.175 & 4.915 & 17.741 & 7.062 & 12.948 & 4.288 & 13.274 &4.375  & 18.531 & 7.265 \\
    &\textbf{Improv.} &\textbf{\footnotesize{$\boldsymbol{\uparrow}$ 7.99\%}} 	&\textbf{\footnotesize{$\boldsymbol{\uparrow}$ 7.46\%}}	&\textbf{\footnotesize{$\boldsymbol{\uparrow}$ 3.69\%}}	&\textbf{\footnotesize{$\boldsymbol{\uparrow}$ 8.55\%}} 	
    & \textbf{\footnotesize{$\boldsymbol{\uparrow}$ 3.27\%}}  &\textbf{\footnotesize{$\boldsymbol{\uparrow}$ 5.50\%}}   	
    & \textbf{\footnotesize{$\boldsymbol{\uparrow}$ 6.96\%}} &\textbf{\footnotesize{$\boldsymbol{\uparrow}$ 3.12\%}}
    & \textbf{\footnotesize{$\boldsymbol{\uparrow}$ 3.32\%}}	& \textbf{\footnotesize{$\boldsymbol{\uparrow}$ 5.66\%}} \\
    \hline 	

    \multirow{3} * {Sports}
    & Base & 7.644 & 2.447 & 10.924 &4.046 &7.939 &2.563 & 7.704 & 2.504 & 11.607 &4.238  \\
    & Ada-Retrieval &8.226 &2.683  & 11.352 &4.234  &8.425 & 2.663  &8.551 & 2.731  &11.903 &4.444  \\
    &\textbf{Improv.} &\textbf{\footnotesize{$\boldsymbol{\uparrow}$ 7.61\%}}  &\textbf{\footnotesize{$\boldsymbol{\uparrow}$ 9.67\%}} 
    & \textbf{\footnotesize{$\boldsymbol{\uparrow}$ 3.92\%}} & \textbf{\footnotesize{$\boldsymbol{\uparrow}$ 4.65\%}} 
    &\textbf{\footnotesize{$\boldsymbol{\uparrow}$ 6.12\%}} &\textbf{\footnotesize{$\boldsymbol{\uparrow}$ 3.90\%}} 
    & \textbf{\footnotesize{$\boldsymbol{\uparrow}$ 11.00\%}}  
    & \textbf{\footnotesize{$\boldsymbol{\uparrow}$ 9.07\%}}  
    &\textbf{\footnotesize{$\boldsymbol{\uparrow}$ 2.55\%}} &\textbf{\footnotesize{$\boldsymbol{\uparrow}$ 4.85\%}}\\
    \hline 
		 	
    \multirow{3} * {Yelp}
    & Base & 9.252 & 2.765 & 12.062 &3.770 &9.828 	&2.971  &10.501  &3.141 & 13.013&4.029  \\
    & Ada-Retrieval &10.415 & 2.985 &12.637 &3.852  & 11.224 &3.316 & 11.688  & 3.454  & 13.430 & 4.157 \\
    &\textbf{Improv.} &\textbf{\footnotesize{$\boldsymbol{\uparrow}$ 12.57\%}}  &\textbf{\footnotesize{$\boldsymbol{\uparrow}$ 7.98\%}} &\textbf{\footnotesize{$\boldsymbol{\uparrow}$ 4.77\%}} &\textbf{\footnotesize{$\boldsymbol{\uparrow}$ 2.19\%}}  
    & \textbf{\footnotesize{$\boldsymbol{\uparrow}$ 14.20\%}} & \textbf{\footnotesize{$\boldsymbol{\uparrow}$ 11.60\%}} &\textbf{\footnotesize{$\boldsymbol{\uparrow}$ 11.31\%}}  & \textbf{\footnotesize{$\boldsymbol{\uparrow}$ 9.96\%}} &\textbf{\footnotesize{$\boldsymbol{\uparrow}$ 3.20\%}} & \textbf{\footnotesize{$\boldsymbol{\uparrow}$ 3.18\%}} \\
    \bottomrule
\end{tabular}
}
 \caption{Top-50 performance comparison of five backbone models and Ada-Retrieval on three datasets. All the metrics in the table are percentage numbers with ’\%’ omitted.}
\label{tab:basemodeltop50}
\end{table*}

We train the base sequential models and their corresponding Ada-Retrieval counterparts using three datasets. The top 50 recommendation results are shown in Table~\ref{tab:basemodeltop50}.

Here, we consistently observe that Ada-Retrieval consistently and significantly outperforms all base sequential models across all datasets and metrics. Notably, Ada-Retrieval (GRU4Rec) exhibits an average improvement of 8.37\% in terms of NDCG@50 over GRU4Rec on the three datasets, while Ada-Retrieval (SASRec) demonstrates an improvement of 5.57\% over SASRec.

Whether it is RNN-based (GRU4Rec), Transformer-based (SASRec), CNN-based (NextItNet), GNN-based (SRGNN), or MLP-based (FMLPRec), Ada-Retrieval seamlessly integrates and consistently enhances performance. Notably, Ada-Retrieval shares the same embedding layer, user model architectures, and prediction layers with its respective base models. This shared architecture underscores its effectiveness in adapting a base model according to varying contextual information for the specific task. In essence, Ada-Retrieval embodies a plug-and-play property, allowing the augmentation of any given base model with adaptation modules while preserving the original architecture's integrity.

\subsection{Comparison with Multi-Interest Models}

\subsubsection{Baselines.}
Our method employs multi-round adaptive learning to progressively generate multiple user representations. Although it fundamentally differs from another research direction, multi-interest-aware user modeling, there are several similarities between the two approaches, such as producing multiple user representations during inference. In this regard, we compared Ada-Retrieval with several multi-interest retrieval models as baselines: \textbf{DNN}~\cite{covington2016deep} (also known as \textsl{YouTube DNN}), \textbf{MIND}~\cite{li2019multi}, \textbf{ComiRec}~\cite{cen2020controllable}, and \textbf{SINE}~\cite{tan2021sparse}

\subsubsection{Results.}

\begin{table}[t!]
\scalebox{0.8}{
    \begin{tabular}{l|cc|cc|cc} 
        \toprule
        \multirow{2} * {Methods} & \multicolumn{2}{c}{Beauty} & \multicolumn{2}{|c}{Sports} & \multicolumn{2}{|c}{Yelp} \\
         & HR & NDCG & HR & NDCG & HR & NDCG \\
        \midrule
        DNN &13.705 &4.726   &8.798 &2.890   & 11.241 &3.317 \\
        \midrule
        MIND &14.045 &5.002  & 8.888 &2.918 &11.320 	&3.443 \\ 
        ComiRec & \underline{14.394}  &\underline{5.232}    &\underline{9.270}  &	\underline{3.250}  & 11.479  & 3.523  \\ 
        SINE & 13.191  &4.325    &9.087  &2.978  & \underline{12.091}  &\underline{3.724}  \\ 
        \midrule
       Ada. &\textbf{17.741} 	&\textbf{7.062}   &\textbf{11.352} &\textbf{4.234} & \textbf{12.637} &\textbf{3.852} \\
        \bottomrule
    \end{tabular}
	}
 \caption{Top-50 performance comparison of several baselines and Ada-Retrieval (SASRec) on three datasets.} 
\label{tab:baselinestop50}
\end{table}

The overall results are presented in Table~\ref{tab:baselinestop50}. It is evident that approaches utilizing multiple user representation vectors (such as MIND, ComiRec, and Ada-Retrieval) exhibit superior performance compared to those employing a single representation (DNN). This finding highlights the effectiveness of multiple user representation vectors in capturing diverse user interests and, consequently, elevating recommendation accuracy.

Generally, Ada-Retrieval consistently outperforms other models across all metrics on the three datasets, underscoring its effectiveness. This success can be attributed to two key factors: 1) Ada-Retrieval's multi-round retrieval paradigm, which refines user representations iteratively based on contextual information, enabling more precise identification of potential candidates across the entire item space. 2) Unlike multi-interest methods that rely on heuristic rules to determine the number of interests, Ada-Retrieval autonomously discerns the depth and range of users' preferences.

\subsection{Ablation Study}

\begin{table}[t!]
\scalebox{0.8}{
    \begin{tabular}{l|cc|cc|cc} 
        \toprule
        \multirow{2} * {Methods} & \multicolumn{2}{c}{Beauty} & \multicolumn{2}{|c}{Sports} & \multicolumn{2}{|c}{Yelp} \\
         & HR & NDCG & HR & NDCG & HR & NDCG \\
        \midrule
        Base & 17.110 &6.506   &10.924 &	4.046   & 12.062 &3.770 \\
        \midrule
          w/o LFT &17.549 &6.945   & 11.287 &	4.186  & 12.474 	&3.834 \\ 
         w/o CAT & 17.536 &6.948   &11.141 &	4.182 & 12.566 & 3.826 \\ 
          w/o IRA & 17.209 &6.670   & 10.980 &	4.101  & 12.331 	& 3.815  \\
        \midrule
        w/o GRU  &17.348 &6.845   & 11.020 &4.137  &12.464 &3.817 \\ 
        w/o MLP & 17.227 &6.716   &10.692 	&3.963   & 12.026 &	3.691  \\
        w/o URA & 17.308 &6.793  &10.782 &	4.017  & 12.147 	&3.720   \\
        \midrule
        w/o PT  &17.200 &6.553   &10.739 	&3.969   & 12.008 &	3.692  \\
        \midrule
       Ada. &\textbf{17.741} 	&\textbf{7.062}   &\textbf{11.352} &\textbf{4.234} & \textbf{12.637} &\textbf{3.852} \\
        \bottomrule
    \end{tabular}
	}
 \caption{Results of Ablation Study.} 
\label{tab:ablation_study}
\end{table}

Our proposed Ada-Retrieval includes a learnable filter layer (LFT), a context-aware attention layer (CAT) within the item representation adapter (IRA), and a GRU layer and MLP layer in the user representation adapter (URA). We conduct an ablation study comparing Ada-Retrieval (SASRec) with SASRec on three datasets to analyze the contribution of each part. Additionally, we explore different training strategies, such as without pre-training (w/o PT) in Ada-Retrieval. The results are reported in Table~\ref{tab:ablation_study}.

Upon omitting the filter layer, there is a discernible drop in performance, suggesting that learnable filters play a pivotal role in mitigating the effects of noisy data within the item context. When we replace the attention layer with the average of item embeddings in context, the decline in performance suggests that the attention mechanism allows the model to automatically identify key items by assigning higher weights to them. The most pronounced degradation is observed when the entire item context information was removed. This highlights the vital role of item context data in Ada-Retrieval's user simulation process.

In terms of user context, removing either the GRU layer or the MLP module results in a significant performance drop compared to Ada-Retrieval, highlighting the effectiveness of our user representation adapter in integrating context information. Notably, omitting the MLP causes a more pronounced decline in performance than using the model without the URA. This suggests that directly incorporating the user context vector into the current user representation introduces noise, emphasizing the importance of carefully designing a fusion module to effectively leverage user context.

Additionally, jointly training $\Phi$ and $\Theta$ from scratch results in inferior performance compared to Ada-Retrieval in three datasets, highlighting the significance of the two-stage training procedure. This can be attributed to the pre-trained base model's ability to generate more accurate and robust context information, which facilitates the training of the model.

\begin{figure}[t]
\centering
\includegraphics[width=1.0\columnwidth]{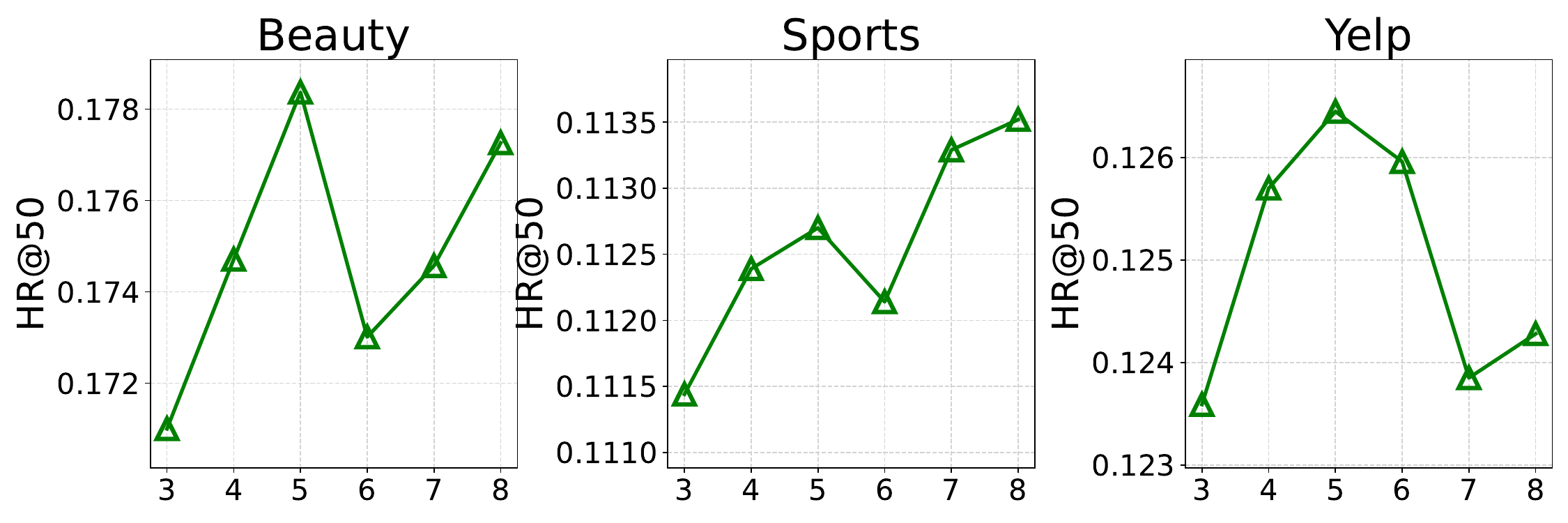} 
\caption{Effect of the number of recommendation batches.}
\label{fig:hyper-T}
\end{figure}

\begin{figure}[t]
\centering
\includegraphics[width=1.0\columnwidth]{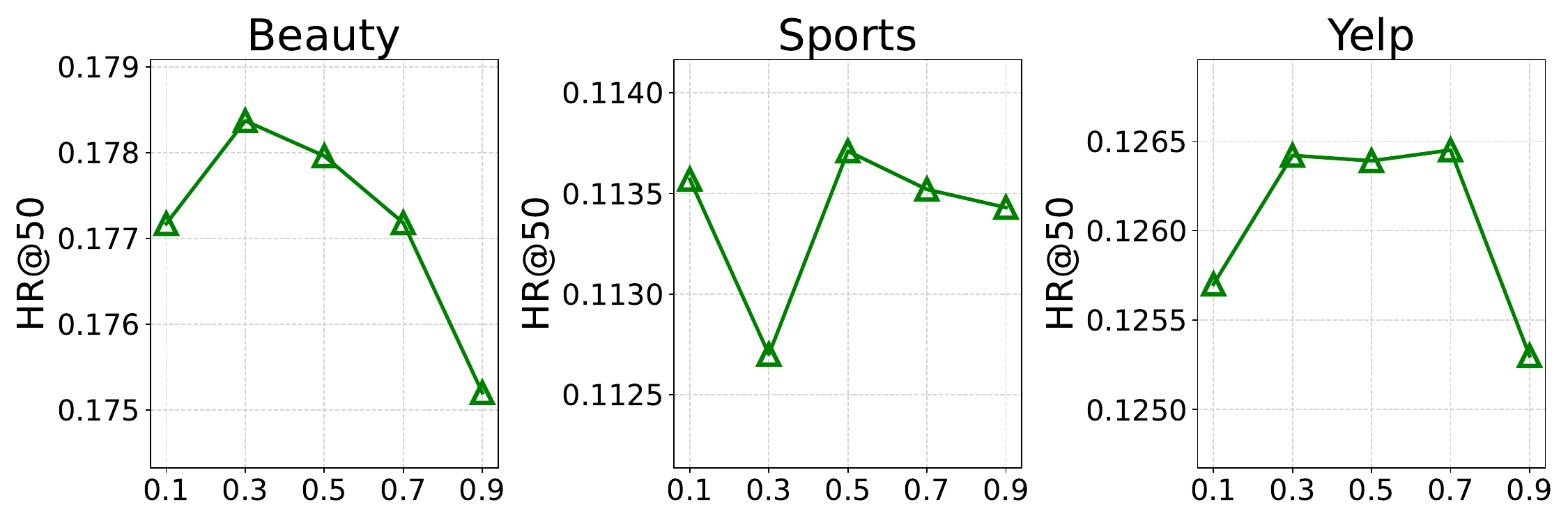} 
\caption{Sensitivity analysis of parameter $\lambda$.}
\label{fig:hyper-lam}
\end{figure}

\subsection{Hyper-parameter Analysis}
We further investigate the impact of our model's hyperparameters, specifically $T$ and $\lambda$, on three datasets.

In Figure~\ref{fig:hyper-T}, the optimal performance is observed when $T$ is set to 5 for the Beauty dataset, 8 for Sports, and 6 for Yelp. The model's performance exhibits a monotonically increasing trend as $T$ rises from 1 to the optimal $T^{\ast}$. However, exceeding $T^{\ast}$ introduces unpredictability due to excessive inference rounds.

Figure~\ref{fig:hyper-lam} reveals that the performance of Ada-Retrieval initially increases with the rise of $\lambda$. It gradually reaches its peak when $\lambda$ is 0.3 for Beauty, 0.5 for Sports, and 0.7 for Yelp. Subsequently, the performance begins to decline. When the factor $\lambda$ is too low or high, it fails to provide useful supervisory information for training. Therefore, choosing an appropriate value for $\lambda$ with a validation set is crucial.

\subsection{Analysis of Each Round}

\begin{figure}[t]
\centering
\includegraphics[width=1.0\columnwidth]{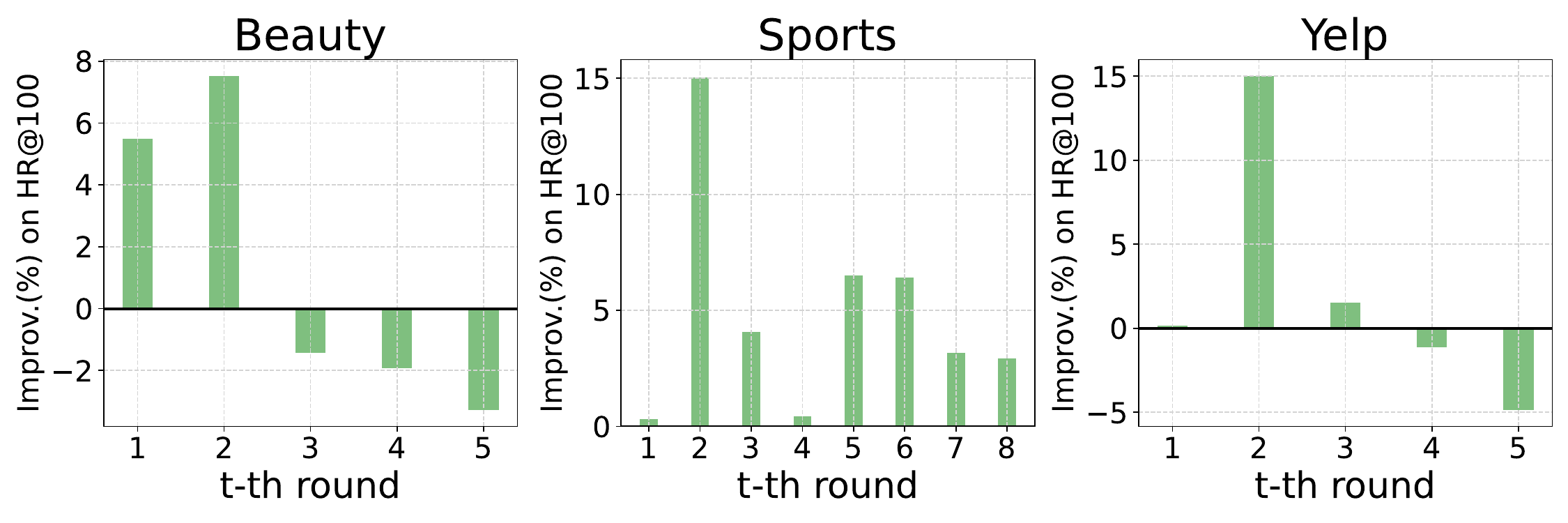} 
\caption{Improvement between Ada-Retrieval and SASRec on each round.}
\label{fig:anal-turn}
\end{figure}

One core aspect of the Ada-Retrieval model is its cascading multi-round preference modeling of users. Thus, we compare the performance difference between Ada-Retrieval and the base model in each turn, assessing the improvement in top-$k$ recommendations made by Ada-Retrieval (SASRec) over SASRec, as depicted in Figure~\ref{fig:anal-turn}.

Ada-Retrieval consistently exhibits substantial improvements in the early rounds, achieving a 7.5\% enhancement on Beauty, 15\% on Sports, and Yelp in terms of HR@100. However, as the rounds progress, the performance advantage narrows to a modest 3\% on Sports and experiences a slight dip of -2.5\% on Beauty. Nevertheless, when considering the overall performance, Ada-Retrieval consistently outperforms. This suggests that Ada-Retrieval excels at rapidly and precisely elevating items—those that the base model either misses or ranks lower—during the preliminary rounds.

\section{Conclusion}
In this paper, we introduce Ada-Retrieval, a novel adaptive multi-round retrieval paradigm for sequential recommendations, which provides a more dynamic and diverse approach compared to the traditional single-round inference paradigm. This model-agnostic framework incorporates key components, such as the item representation adapter and user representation adapter, to effectively refine the retrieval process in a progressive manner. Extensive experiments on publicly available datasets demonstrate the effectiveness of Ada-Retrieval, emphasizing its potential to enhance the performance of various sequential recommender systems. Future work may include investigating the theoretical foundations of the benefits provided by the multi-round retrieval paradigm and expanding its application to large-language
models, such as augmenting task-planning abilities.

\section*{Acknowledgement}
This work was supported by the National Natural Science Foundation of China (NSFC Grant No.62106274); the Fundamental Research Funds for the Central Universities, Renmin University of China (No.22XNKJ24). We also wish to acknowledge the support provided by the Intelligent Social Governance Platform, Major Innovation \& Planning Interdisciplinary Platform for the "Double-First Class" Initiative.

\bibliography{aaai24}

\end{document}